\documentclass[11pt]{article}

\usepackage[utf8]{inputenc}
\usepackage[T1]{fontenc}
\usepackage[USenglish]{babel}
\usepackage[nodayofweek,level]{datetime}
\usepackage[pdftex]{graphicx}
\usepackage[hyperfootnotes=false]{hyperref}
\usepackage{pdfpages}
\usepackage{listings}

\usepackage{amsmath}
\usepackage{amsfonts}
\usepackage{amsthm}
\usepackage{subfig}

\usepackage{tabularx}
\usepackage{booktabs}

\usepackage{authblk}

\usepackage{lineno}
\usepackage{geometry}
\makeatletter
\def\blfootnote{\xdef\@thefnmark{}\@footnotetext}
\makeatother

\title{Migration patterns under different scenarios of sea level rise}
\author[1]{Caleb Robinson}
\author[2]{Bistra Dilkina\footnote{Corresponding author: \texttt{dilkina@usc.edu}}}
\author[3]{Juan Moreno-Cruz}
\affil[1]{\small Georgia Institute of Technology, School of Computational Science and Engineering, Atlanta, Georgia.}
\affil[2]{\small University of Southern California, Viterbi School of Engineering, Los Angeles, California.}
\affil[3]{\small University of Waterloo, School of Environment, Enterprise and Development, Waterloo, Ontario.}
\date{\today}

\begin{document}

\maketitle
\blfootnote{\textbf{Author contributions} C.R., B.D. and J.M-C designed research. C.R. performed research. C.R., B.D. and J.M-C analyzed results. C.R., B.D. and J.M-C wrote the paper.}

\begin{abstract}
We propose a framework to examine future migration patterns of people under different sea level rise scenarios using models of human migration. Specifically, we couple a sea level rise model with a data-driven model of human migration, creating a generalized joint model of climate driven migration that can be used to simulate population distributions under potential future sea level rise scenarios. We show how this joint model relaxes assumptions in existing efforts to model climate driven human migration, and use it to simulate how migration, driven by sea level rise, differs from baseline migration patterns. Our results show that the effects of sea level rise are pervasive, expanding beyond coastal areas via increased migration, and disproportionately affecting some areas of the United States. The code for reproducing this study is available at \url{https://github.com/calebrob6/migration-slr}.
\end{abstract}

%--------------------------------------------------------------------
%--------------------------------------------------------------------
%--------------------------------------------------------------------
\section*{Introduction}
Sea level rise (SLR) will affect millions of people living in coastal areas.
According to the IPCC Fifth Assessment Report, in the ``worst-case'' Representative Concentration Pathways (RCP) scenario, RCP 8.5, where greenhouse gas emissions continue to rise throughout the 21st century, a global mean sea level (GMSL) rise between 0.52 to 0.98 meters (m) is \textit{likely} by 2100~\cite{church01ipcc}. Other estimates, using statistical instead of process based models of GMSL, project a rise in the range of 0.75 m to 1.9 m by 2100~\cite{vermeer2009global}. Recent research from the National Oceanic and Atmospheric Administration (NOAA), however, has suggested a 2.5 m upper bound of GMSL rise by 2100 for an `extreme' sea level rise scenario, and a 2 m GMSL rise for a `high' scenario~\cite{sweet2017global}.

The impacts of sea level rise are potentially catastrophic. About 30\% of the urban land on earth was located in high-frequency flood zones in 2000, and it is projected to increase to 40\% by 2030 taking urban growth and sea level rise into account~\cite{guneralp2015changing}.
In the United States alone, 123.3 million people, or 39\% of the total population, lived in coastal counties in 2010, with a predicted 8\% increase by the year 2020~\cite{crossett2014national}. By the year 2100, a projected 13.1 million people in the United States alone would be living on land that will be considered flooded with a SLR of 6 feet (1.8 m)~\cite{hauer2016millions}.

Human migration is a natural response to this climatic pressure, and is one of many adaptation measures that people will take in response to climate change~\cite{mcleman2006migration, feng2010linkages, wilby2012adapting,kahn2013climatopolis}. 
As oceans expand and encroach into previously habitable land, affected people - climate migrants - will move towards locations further inland, looking for food and shelter in areas that are less susceptible to increased flooding or extreme weather events. In this paper, we argue that the comprehensive impacts of sea level rise on human populations, when considering migration, expand far beyond the coastal areas.

Sea level rise impacts are a combination of two effects.
The \textit{direct} effects of SLR capture the amount of land that will be flooded and the number of people that will be displaced and forced to relocate as an ultimate result of the loss of habitable land.
The \textit{indirect} effects of SLR are more nuanced and are the results of aggregate changes in population distributions across the land.
These indirect effects will cause accelerated changes for inland areas, particularly urban areas, that will observe much higher levels of incoming migrants than they would have absent SLR. These changes can in turn take the form of tighter labor markets~\cite{borjas2016labor} and increased housing prices~\cite{business2014economic}, with broader effects on income inequality in the coastal areas~\cite{shayegh2017outward}. Of course, migration to other cities can also have positive impacts; new migrants can improve productivity as they bring with them human capital accumulated elsewhere~\cite{kahn2013climatopolis}.

Discussions regarding sea level rise impacts on human populations are often constrained to regions \textit{directly} experiencing SLR-driven flooding~\cite{nicholls2002analysis,willekens2016international,hauer2016millions}. Several theoretical frameworks use qualitative case studies to motivate models that might represent the reasoning behind migration choices due to sea level rise, but are not grounded in statistical methods~\cite{mcleman2006migration,black2011effect,piguet2010linking}. There are many complex interactions between demographic driven migration and climate change driven migration, and the scope and scale of the impacts of climate change on migration will be significant~\cite{feng2010linkages, hugo2011future}. One example of these impacts that has been studied considers the political ramifications that will come with the eventual migrants from Pacific island of Kiribati, which will most likely become completely flooded under a 3 meter SLR in the coming centuries~\cite{wyett2014escaping}. Another example is the projected widening demographic differentials in countries that will be especially impacted by sea level rise~\cite{hugo2011future, curtis2011understanding}, similar to demographic consequences seen after the 1970s droughts in Africa~\cite{caldwell1975sahelian,adamo2012impact}. The foundation of both of these concerns is in people's destination locations, therefore it is prudent to weigh the question of `where' people will go equally with `how many' people will be initially affected~\cite{findlay2011migrant}.

There are few empirical studies that link climate change with human migration patterns. Feng et al. show that the negative impacts of climate change on crop-yields has driven increased emigration from Mexico to the United States~\cite{feng2010linkages}, while Thiede and Gray examine the effects of changing climate variables on the timing of migration in Indonesia~\cite{thiede2017heterogeneous}. The only empirical works that examine the \textit{effects of SLR on human migration} do so by coupling population projections with sea level rise and migration models to estimate how population distributions might change in future scenarios~\cite{hauer2017migration, davis2018universal}. In the US, small area population projections for the year 2100 have been combined with spatially explicit estimates of SLR~\cite{hauer2016millions} and an unobserved component regression model to estimate the destinations of populations that could be forced to migrate through coastal flooding. In \cite{hauer2017migration}, approximately 56\% of counties in the US are found to be affected by larger migrant influxes under 1.8 m of SLR. Similarly, in Bangladesh, gridded population projections have been combined with a \textit{bathtub} type model of SLR and the radiation model of human migration to estimate how population distributions may change~\cite{davis2018universal}. This coupled model has minimal data requirements, forecasts large quantities of immigrations to the division of Dhaka in Bangladesh, and highlights the broader potential impacts of these migrants including an increased demand for housing, food, and jobs. 

These empirical studies make the critical simplifying assumption that climate driven migration will follow the same patterns as historic migration. Additionally, ~\cite{hauer2017migration} assumes that migrations will happen only between locations for which there are historically observed migrations. However, human migration is a function of push and pull factors, and where increased climate stress will affect both~\cite{black2011effect}. As such, the patterns of climate migrants will not necessarily follow patterns observed in historical migration data. Indeed, ``climate migrants resulting from press stressors will probably constitute `enhanced', or extra, normal out-migration.''~\cite{hauer2017migration}.

In this work we aim to address the simplifying assumptions made in previous empirical analyzes. First, we \textbf{derive a general framework for coupling models of sea level rise with dynamic models of human migration} so that future innovations in both human migration modeling and sea level rise modeling can easily be coupled to produce more precise estimates of changing population distributions. Second, we \textbf{design our framework with separate human migration models for affected and unaffected populations} in order to capture the different dynamics of these processes. Specifically we implement our framework with small area population projections~\cite{hauer2016millions}, the NOAA's Digital Coast Sea Level Rise estimates~\cite{marcy2011new}, and a recent machine learning (ML) method for modeling human migration~\cite{robinson2018machine}. We model migrations made from \textit{flooded} areas and from \textit{unflooded} areas separately by fitting one ML migration model using ``business-as-usual'' migration data and one model with migration data following Hurricanes Katrina and Rita. Furthermore, our general framework also separates flooded and unflooded zones at a conceptual level to ensure simulated migrations will not end in flooded areas.

In our analysis, we compare simulated migration patterns under two sea level rise scenarios (1.8m SLR in 2100, and 0.9m SLR in 2100) to the baseline scenario of no SLR. We examine the aforementioned \textit{direct} and \textit{indirect} effects of SLR through its impact on human migration, and show results that highlight the importance of treating the dynamics of climate-induced migrants separately from business-as-usual migrations. Our analysis aims to answer the question of how the population distribution will change under different amounts of SLR, i.e. ``where'' populations displaced by SLR will go, and further, how SLR will affect the broader dynamics of human migration.

%--------------------------------------------------------------------
%--------------------------------------------------------------------
%--------------------------------------------------------------------
\section*{Modeling Framework}

%--------------------------------------------------------------------
%--------------------------------------------------------------------
%--------------------------------------------------------------------
\subsection*{Conceptual Challenges}
Current state-of-the-art models of human migration include the family of radiation models~\cite{simini2012universal,yang2014limits}, gravity models~\cite{lenormand2012universal,lenormand2016systematic}, and machine learning models~\cite{robinson2018machine}. Yet, SLR-driven human migration poses some specific challenges to traditional strategies for modelling human migration:

First, human migration induced by SLR might not follow historic migration patterns. In fact, using one year of county-to-county migration data from the IRS U.S. migration data sets, Simini et al.~\cite{simini2012universal} showed that a fixed proportion (~3\%) of the population of a U.S. county will migrate under normal circumstances. This will not hold under sea level rise, as the entire population in flooded areas will have to move or adapt in other ways. Importantly, in addition to direct inundation due to sea level rise, climate migrants will be forced to move as climate change effects become more pronounced, directly through the exposure to ``high-magnitude events'' such as large scale flooding from hurricanes, or indirectly through the ``cumulative contribution of ongoing localized events across regions''~\cite{mcleman2013developments}. The dynamics of these environmentally induced migrations will not necessarily follow those of previously observed migrations. We expect that as social scientists gather more data and knowledge, more refined and informed models of human migration will emerge.

Second, the spatial resolution of the migration model must be carefully considered. Climate migrants will not necessarily move large distances as they adapt to changing conditions in inundated areas. Indeed most migrations are made to nearby locations. This phenomenon can be seen in the migrations following Hurricane Katrina in the US, where a majority of destinations were within Louisiana~\cite{fussell2014recovery}. Joint SLR/migration models must capture the situations whereby migrants can choose to move to the unflooded portions of partially flooded zones, and must guarantee that no migrants are assigned to a flooded area.

Finally, SLR will not happen instantaneously and population projections need to account for the cumulative effects of SLR induced migration. The SLR influenced population distribution of a location will diverge from that of a ``business as usual'' scenario as the indirect effects of SLR compound; as more climate migrants settle inland, they will change the migrations patterns of future migrants and so on. Current population projections account for the \textit{direct} effects of SLR where projected populations are made with respect to potentially flooded land~\cite{hauer2016millions}, however these \textit{indirect} effects must also be considered.

%--------------------------------------------------------------------
%--------------------------------------------------------------------
%--------------------------------------------------------------------
\subsection*{General Modeling Framework}

\begin{figure}[t]
\centering
\includegraphics[width=\linewidth]{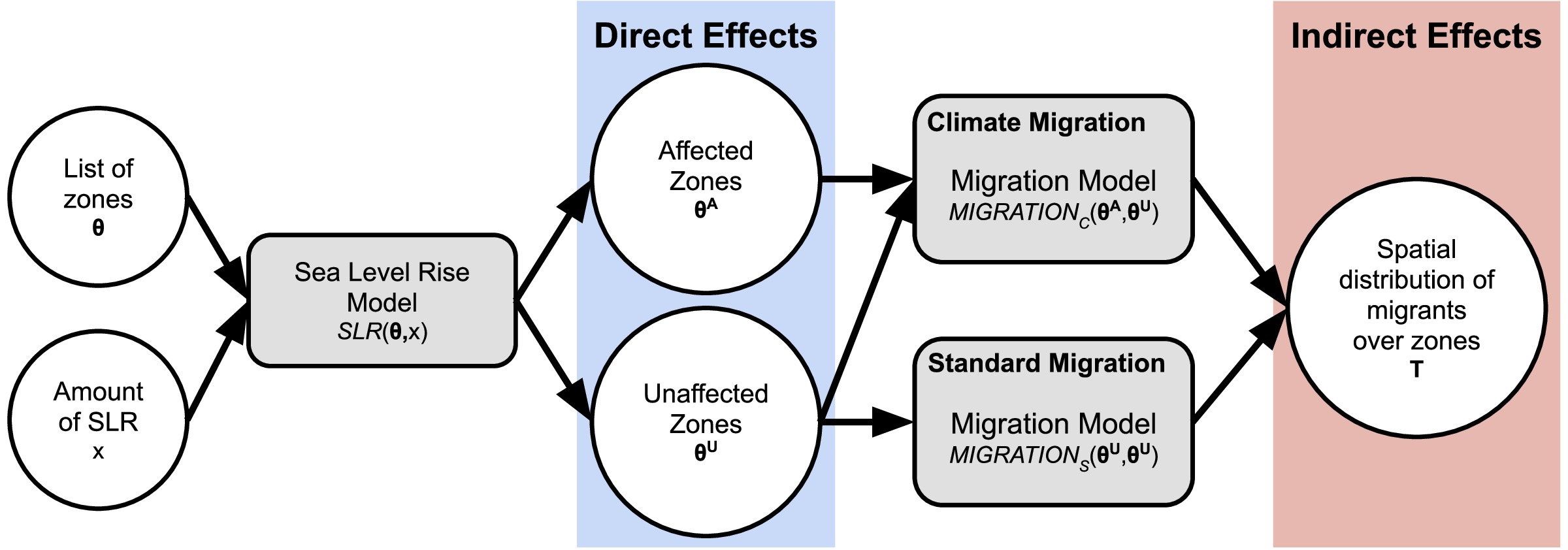}
\caption{\textbf{Joint sea level rise and human migration modeling process.} The joint model takes a list of spatial zones $\boldsymbol{\theta}$ and amount of SLR $x$ as input, and outputs a \textit{migration matrix} $\boldsymbol{T}$, where an entry $T_{ij}$ represents the number of migrants that travel from zone $i$ to $j$ under the given amount of SLR.}
\label{fig:flowChart}
\end{figure}

Consider an amount of SLR in meters, $x$, a list of $n$ spatial zones,  and $\boldsymbol{\theta} = [\theta_1, \ldots, \theta_n]$, which includes the spatial distribution of population in each zone, and, optionally, other features associated with each zone. Using this information, we want to compute a \textit{migration matrix} $\boldsymbol{T}$, where an entry $T_{ij}$ represents the number of migrants that travel from zone $i$ to $j$ under the given amount of SLR. We propose a general modeling framework for handling this problem which consists of two modules, shown in Fig \ref{fig:flowChart}: a \texttt{SLR} module and a \texttt{MIGRATION} module.

\textbf{\texttt{SLR} module}. This module uses a sea level rise model $SLR(\boldsymbol{\theta},x)$ to partition each input zone into two new zones: the flooded portion and the unflooded portion. Using the best available data, this module should divide the features from the original zone ($\theta_i$) between the flooded-portion zone ($\theta^A_i$) and the unflooded-portion zone ($\theta^U_i$). For example, if we have high-resolution spatial population data, then we can split population between the two partitions based on the spatial extent of the flooding. Using this module, we estimate the \textit{direct} effects of SLR in terms of flooded area and number of people living in such an area.

\textbf{\texttt{MIGRATION} module}. This module calculates $\boldsymbol{T}$ using the two sets of zones from the \texttt{SLR} module. Specifically, this module uses two migration models: 1.) a model for migrations from flooded zones $\boldsymbol{\theta}^A$ to unflooded zones $\boldsymbol{\theta}^U$ with the function $MIGRATION_C(\boldsymbol{\theta}^A,\boldsymbol{\theta}^U) = \mathbf{T'}$, where migration is a forced process; and 2.) a model of migrations from unflooded zones to unflooded zones with the function $MIGRATION_S(\boldsymbol{\theta}^U,\boldsymbol{\theta}^U) = \mathbf{T''}$, where migration happens due to usual drivers. Finally, this module should aggregate migrant flows from the two migration functions, $\mathbf{T} = \mathbf{T'} + \mathbf{T''}$. Using this module, we estimate the \textit{indirect} effects of SLR in terms of how the population distribution changes relative to no SLR.

By separating SLR driven migration from standard migrations, our framework forces these dynamics to be considered independently, explicitly bringing up the issue from the first conceptual challenge mentioned in the previous section. Implementations of our framework can use different models for these dynamics if available, or, if such models are not available, fall back to using a simpler model where the simplifying assumption is clear.
Our framework also addresses the second conceptual challenge by excluding flooded areas as destinations for all migrants by simply not considering migrant flows with destinations in the set of flooded portions. Note that within-zone migrations in partially flooded zones are handled by presenting the neighboring unflooded portions of those zones as possible destinations. Furthermore, by separating the functionality of the \texttt{SLR} module from that of the \texttt{MIGRATION} module, the framework allows ablation studies to measure how much results depend on the specific behaviors of each.
The third conceptual challenge revolves around how SLR and migration are both temporal processes that form a feedback loop (i.e. SLR will affect migration decisions, which will in turn affect how many people are affected by further SLR, etc.). This is not explicitly taken into account in our framework, however should be addressed in further research as each process is further understood. In Section 5 of the Supplementary Information we provide a more formal definition of our Joint Model.

%--------------------------------------------------------------------
%--------------------------------------------------------------------
%--------------------------------------------------------------------
\subsection*{Implementation of Joint SLR/Migration Model}

We implement our proposed framework with 1.) small area population projections for 2100 under different amounts of SLR, following the methodology in~\cite{hauer2016millions} - which uses the NOAA's Digital Coast Sea Level Rise estimates~\cite{marcy2011new} - and 2.) a recent machine learning (ML) approach for modeling human migration~\cite{robinson2018machine}. An implementation of the Joint Model requires us to define the \texttt{SLR} function, the manner in which the \texttt{SLR} function splits features associated with the zones that are affected by SLR, and the two \texttt{MIGRATION} functions. All three of these steps are discussed in the next two sections.  In Section 2 of the Supplementary Information we show a similar implementation using the Extended Radiation Model~\cite{yang2014limits} to implement the \texttt{MIGRATION} functions. In Section 3 of the Supplementary Information we describe our ML model and give validation results comparing it to other human migration models on the task of predicting inter-county migrations in the US.

%--------------------------------------------------------------------
%--------------------------------------------------------------------
%--------------------------------------------------------------------
\subsubsection*{Sea Level Rise Modeling}

First, we follow the methodology proposed in Hauer 2016~\cite{hauer2016millions} to create population projections for \textit{every} Census Block Group in the US ($n=216,330$) for two SLR scenarios, \textbf{medium}, where 0.9m (3ft) of SLR is experienced by 2100, and \textbf{high}, where 1.8m (6ft) of SLR is experienced by 2100. These SLR scenarios are also proposed in Hauer 2016, based on methods from the US National Climate Assessment, and use the NOAA's Digital Coast spatial estimates of areas affected by SLR in 1ft (0.3m) increments~\cite{marcy2011new,digitalCoastReport}. The medium SLR scenario expects SLR to reach the 0.3m, 0.6m, and 0.9m thresholds in the years 2055, 2080, and 2100 respectively. The high scenario reaches similar increasing 0.3m SLR increments in the years 2042, 2059, 2071, 2082, 2091, and 2100. The Digital Coast model provides SLR estimates that address many of the shortcomings of a naive \textit{bathtub} calculation of SLR (i.e. thresholding a digital elevation map with expected SLR amounts) by taking tidal variability, hydroconnectivity, probable flooding, and federal leveed areas into account.

With this data, we define the $SLR(\theta_i,x_t)$ function as taking an input \textit{county}, $\theta_i$, and amount of SLR under either the medium or high SLR scenario, $x_t$. A given county corresponds to a set of census block groups while a given SLR amount, under either scenario, corresponds to population projections for the census block groups, including an estimate of the number of people \textit{affected} by flooding in each block group. We split the county population into the affected and unaffected block groups, which directly results in $\theta^U_i$ and $\theta^A_i$. Here, $\theta^U_i$ is a ``new'' county equivalent zone, for the purposes of modeling migration, with a population equal to the sum of the unaffected populations over all the associated block groups. Similarly, $\theta^A_i$, represents the innundated portion of the original county, and will contain a population equal to the sum of affected populations in the associated block groups. We can run $SLR(\theta_i,x_t)$ for all $i$ to get the $\boldsymbol{\theta}^U$ and $\boldsymbol{\theta}^A$ sets. Population is the only feature required by the \texttt{MIGRATION} module, however a similar splitting technique could be used for other block group level features if they can be reliably projected into future scenarios.

\begin{figure}[ph]
\centering
\includegraphics[width=0.85\linewidth]{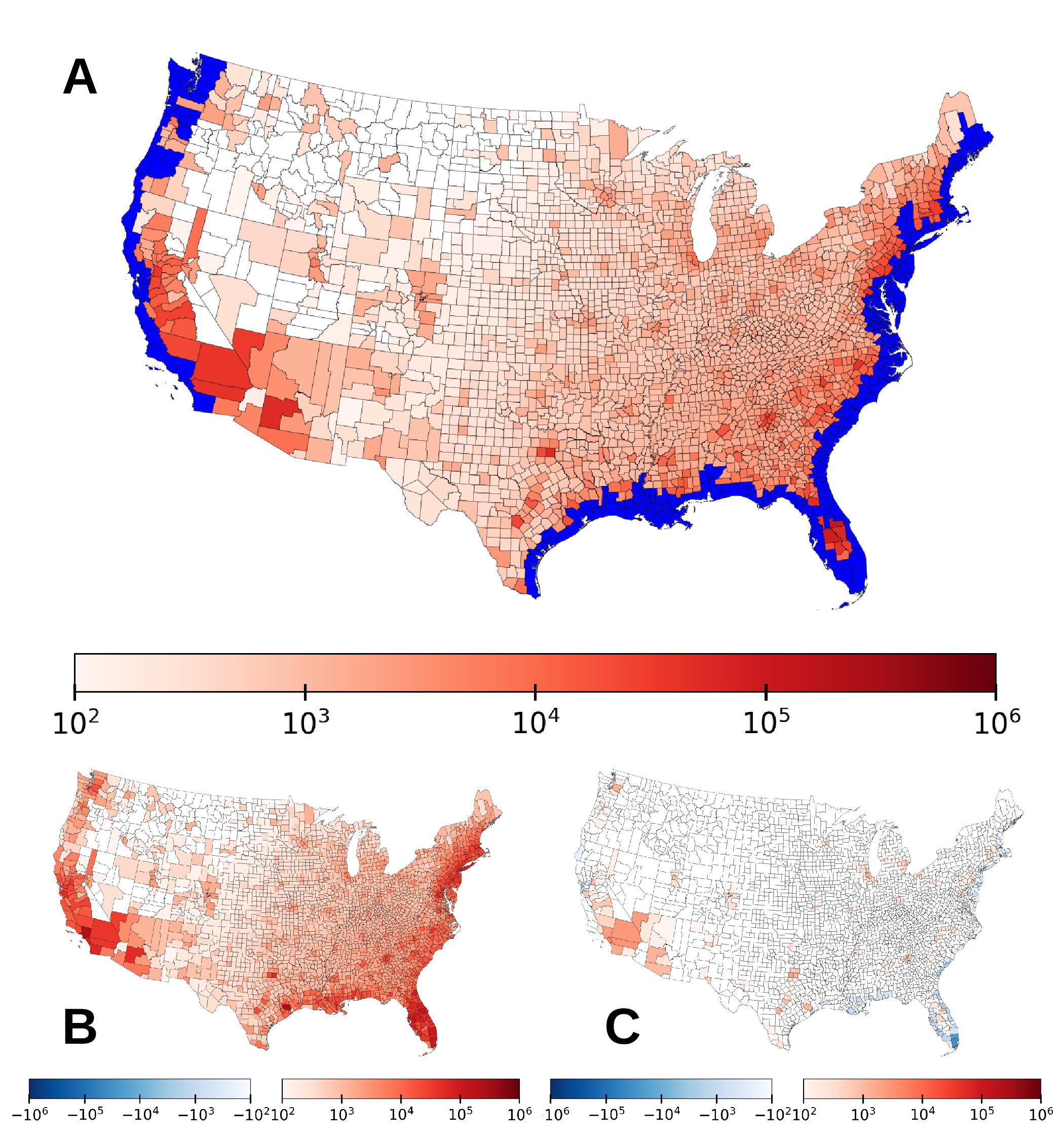}
\caption{\textbf{Spatial distribution of the direct and indirect effects of SLR on human migration.} The \textbf{top} panel shows all counties that experience flooding under 1.8m of SLR by 2100 in blue and colors the remaining counties based on the number of additional incoming migrants per county that there are in the SLR scenario over the baseline. The \textbf{bottom left} map shows the number of additional incoming migrants per county in the SLR scenario from only flooded counties. The \textbf{bottom right} map shows the number of additional incoming migrants per county in the SLR scenario from only unflooded counties. Color gradients are implemented in a log scale.}
\label{fig:2}
\end{figure}

%--------------------------------------------------------------------
%--------------------------------------------------------------------
%--------------------------------------------------------------------
\subsubsection*{Human Migration Modeling}

We model human migration between counties in the USA with a recently proposed artificial neural network (ANN) based method~\cite{robinson2018machine} that is fit with historic county-to-county migration data from the IRS~\cite{irsData}. This method is similar in functionality to traditional models of human mobility and migration, such as the radiation or gravity models~\cite{simini2012universal,erlander1990gravity,lenormand2012universal}, as it will estimate the probability of a migration between a given origin and destination based on population and distance features~\cite{simini2012universal}.

More specifically, our ANN models estimate $P_{ij}$, the probability that a migrant which leaves an origin county, $i$, will travel to a destination county, $j$, using the following input features: origin population, $m_i$, destination population, $m_j$, distance between the two, $d_{ij}$, and the ``intervening opprtunities'' between the two, $s_{ij}$ (this is the total population in the circle centered at $i$ with radius $d_{ij}$, not including $m_i$ or $m_j$). These features are the same features used by traditional radiation and gravity models, and depend solely on population and distance.

To compute $T_{ij}$, the number of migrants that travel from $i$ to $j$, we need to know the number of migrants that are attempting to leave $i$. If we say that the number of migrants leaving zone $i$ is of the form $g(m_i) = \alpha m_i$, where $\alpha$ is some coefficient that specifies the fraction of the total population that will migrate, then $T_{ij} = g(m_i) P_{ij}$. This function $g$ is called the production function. Now we can define the \texttt{MIGRATION} functions, $MIGRATION_C$ and $MIGRATION_S$, which represent the climate migrants and standard migrants respectively, by training two instances of our ANN model, and forming respective production functions $g_C(m_i)$ and $g_S(m_i)$ by choosing $\alpha_C$ and $\alpha_S$.

We fit the $MIGRATION_C$ model by finding hurricane affected counties from the IRS migration data from 2004-2011 and 2011-2014. Specifically, we search for migration data points (i.e. pairs of counties) in which the origin county was a coastal county that observed an over 100\% \textit{increase} in outgoing migrations with over 1,000 total outgoing migrations\footnote{The reporting methodology in the IRS migration dataset changed between the 2010-2011 data and 2011-2012 data, therefore we cannot measure percent increase in outgoing migrations between them.}. This ``filter'' highlights counties that have potentially been affected by hurricanes or other natural disasters and indeed finds seven counties from 2005 that were heavily impacted by hurricanes Katrina and Rita: St Bernard, Orleans, Cameron, Plaquemines, Hancock, Jefferson, and Harrison (which matches literature estimates of the most damaged counties~\cite{fussell2014recovery}), as well as Liberty County, GA from 2006. The only historical explanation we can find for a sudden increase in outgoing migration in Liberty County is the deactivation of a large US military division stationed within the county that year. Considering this, we fit our $MIGRATION_C$ model using the data from the seven counties that were most seriously affected by hurricanes Katrina and Rita. As background, Hurricane Katrina struck the city of New Orleans and the broader Louisiana and Mississippi coastline on August 29, 2005 causing widespread flooding and wind damage. Less than a month later, on September 25, Hurricane Rita also struck the Louisiana coast, exacerbating damage in New Orleans and causing extensive damage to counties in the western portions of the state. Over 1500 people were killed and over 80\% of the city of New Orleans was flooded as a result of these hurricanes. Followup studies and Census estimates showed that New Orleans only contained around half of its pre-hurricane population within a year of the storms. By training our ANN with these counties we allow the model to pick up on the dynamics of migrations after extreme flooding events.

We train an ANN, $MIGRATION_C(\theta^A_i, \theta^U_j) = P'_{ij}$, using all $7 \times 3099$ pairs of counties from the 2005-2006 IRS data that include one of the seven previously mentioned \textit{affected} counties as an origin and an \textit{unaffected} county as a destination. Similarly, we fit another ANN using the rest of the IRS migration data, $MIGRATION_S(\theta^U_i, \theta^U_j) = P''_{ij}$. Due to our assumption that all people in flooded areas will have to migrate, the production function for climate migrants is given as the identity, $g_C(m_i) = m_i$. This forces the entire population of the affected portions of counties to become migrants. For standard migrants, we use the production function, $g_S(m_i) = 0.03 m_i$, due to the observation by Simini et al.~\cite{simini2012universal} that 3\% of a county's population will migrate under normal conditions each year. Given these: $T'_{ij} = g_C(m_i) MIGRATION_C(\theta^A_i, \theta^U_j)$, and $T''_{ij} = g_S(m_i) MIGRATION_S(\theta^U_i, \theta^U_j)$. With these definitions we can build $\mathbf{T'}$ and $\mathbf{T''}$ by running the climate migration ANN and standard migration ANN for all pairs of counties. In Section 3 of the Supplementary Information we evaluate the performance of these models compared to other migration models in a cross validation experimental setup.

%--------------------------------------------------------------------
%--------------------------------------------------------------------
%--------------------------------------------------------------------
\section*{Results}

\begin{figure}[ph]
\centering
\includegraphics[width=0.85\linewidth]{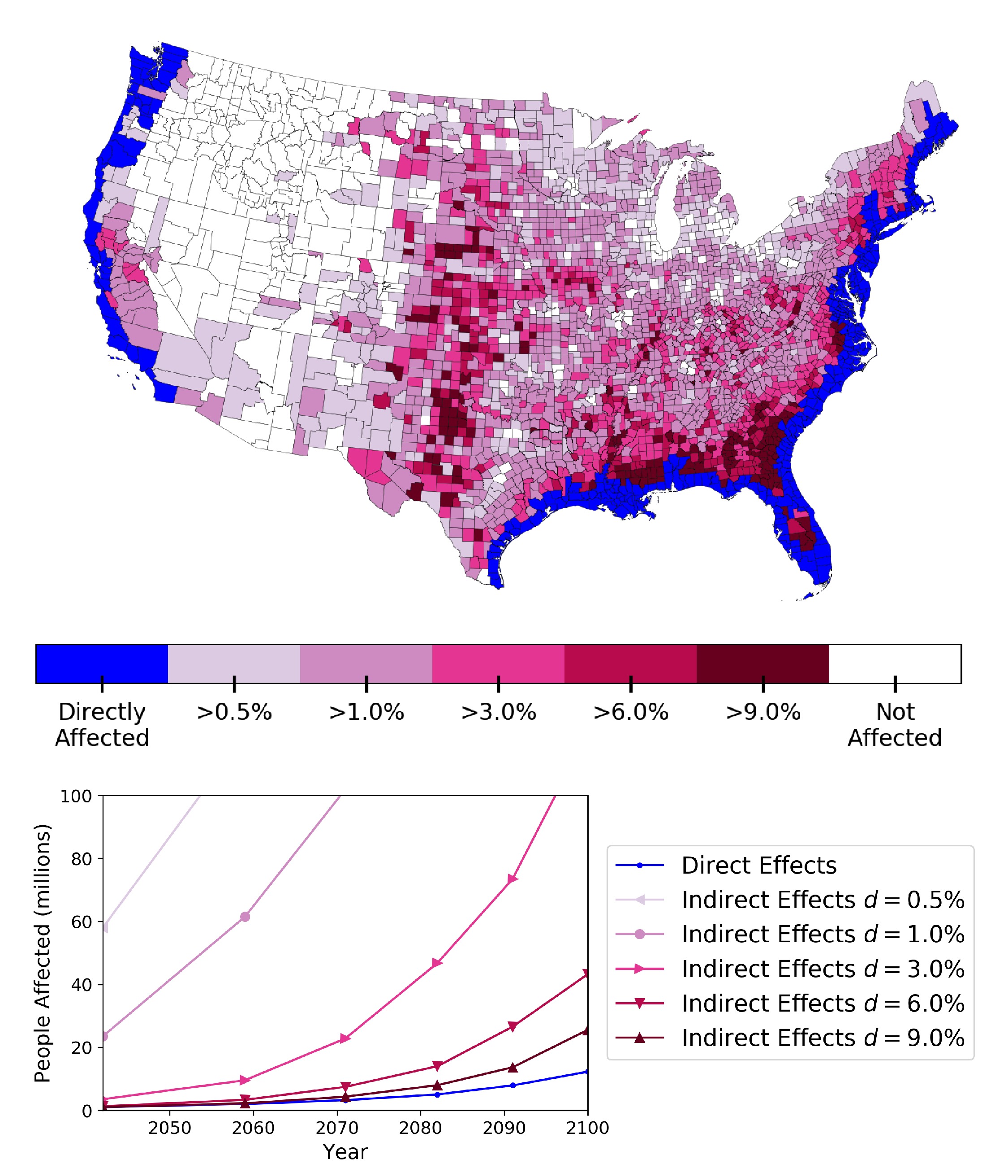}
\caption{\textbf{Impacts of SLR due to flooding and human migration for a range of SLR scenarios.}  We say that a county is indirectly affected by SLR if the difference between the number of incoming migrants to the county in the SLR scenario and the number of incoming migrants in the baseline scenario, i.e. the number of \textit{extra} migrants in the SLR scenario, is greater than some percentage, $d$, of that county's population. In the \textbf{top} panel we show the spatial distribution of counties that are considered indirectly affected at different threshold values of $d$ for the 1.8m SLR case in the southeast portion of the United States. In the \textbf{bottom} panel we show the number of people that are directly and indirectly affected under the same threshold values of $d$ for the entire United States. For both plots we show aggregate impacts for five different values of $d$: 0.5\%, 1\%, 3\%, 6\%, and 9\%.}
\label{fig:3}
\end{figure}

We categorize the effects of SLR into two types: \textit{direct effects}, which are a direct consequence of SLR, and \textit{indirect effects}, which are a consequence of changing migration patterns due to SLR. We present the spatial distribution and magnitude of these effects in Figs \ref{fig:2} and \ref{fig:3}. People that live on flooded land who will have to move away are accounted for in the \textit{direct effects} of SLR. People that live in counties that experience a larger number of incoming migrants in the flooding scenario relative to the baseline scenario with no SLR are accounted for in the \textit{indirect effects} of SLR.

In Fig \ref{fig:2} we show the spatial distribution of changes in migration patterns. In the top panel, counties experiencing any flooding (i.e. that are directly affected by SLR) are highlighted in blue, while the remaining counties are colored according to how many additional migrants they receive in the 1.8m SLR scenario. The bottom two panels of Fig \ref{fig:2} show the difference in the number of incoming migrants between the SLR scenario and the baseline scenario for incoming migrants from \textit{unaffected} counties and \textit{affected} counties. The top panel is the sum of these two maps, and shows this difference for incoming migrants from \textit{all} counties.  From these maps we can see that the primary destination of climate migrants are counties just inland of their origin, but climate migrants also move farther towards large cities that offer more opportunities.

In Fig \ref{fig:3} we show the magnitude of the \textit{direct} and \textit{indirect} effects as well as the spatial distribution of the indirectly affected counties. Formally, a county is marked as indirectly affected if the difference between the number of incoming migrants in the SLR scenario and the baseline scenario is greater than a percentage $d\%$ of the population of that county. By varying $d$ we can see different intensities of indirect effects. On average, 3\% of a county's population migrates each year~\cite{simini2012universal}. Thus, if $d=3\%$, we would observe twice as much migration into a particular county than the average migration rate of the US. We assume that as $d$ increases, the effects will be stronger as there will be a significant strain on the resources in that particular county. 

The graph in the bottom panel of Fig \ref{fig:3} shows the direct and indirect effects of SLR in terms of number of people affected for amounts of SLR in the range from 0.3m to 1.8m in 0.3m increments. The map in the top panel shows which counties in the United States are indirectly affected at different threshold values of $d$. In both plots the indirect effects are shown for five different values of $d$: 0.5\% 1\%, 3\%, 6\%, and 9\%.

From the graph in Fig \ref{fig:3} we can see that the \textit{indirect} impacts of SLR grow at much faster rate than the \textit{direct} impacts. In the high SLR scenario by the year 2100 there are $\approx$ 13 million people \textit{directly} affected, in $\approx$ 50 thousand km$^2$ of flooded land, however there are almost twice as many, $\approx$ 25 million people, \textit{indirectly} affected at the 9\% threshold due to changing migration patterns and magnitudes. This 9\% threshold indicates that these people live in areas which will experience three times as many migrants as they would compared to a baseline scenario. Even under the moderate assumption of 0.9m SLR by 2100 there will be 24 million people that live in counties considered indirectly affected at a 3\% threshold. Under the same threshold with a SLR of 1.8m by 2100, there will be 120 million people, over $\approx 1/3$ of the population of US, living in counties that will see a doubling in the number of annual incoming migrants.

The map in Fig \ref{fig:3} shows that these \textit{indirect} effects relative to county population will be distributed unevenly over the US. Most effects are seen in the Eastern US, where there are more vulnerable coastal populations. Of particular note are southern Mississippi and southeastern Georgia, where large groups of counties are estimated to see indirect effects in the $>9\%$ category. The Midwest is also projected to see large indirect effects, even though the magnitudes of incoming migrants are smaller than counties closer to the coast. This can be explained by the relatively small populations and baseline levels of incoming migrants. The greater magnitudes of migrations from higher population areas causes \textit{some} migrants to select these midwestern areas as destination, which could cause disproportionally larger \textit{indirect} effects.

%--------------------------------------------------------------------
%--------------------------------------------------------------------
%--------------------------------------------------------------------
\section*{Discussion}

Our results show that the effect of SLR on human populations could be more pervasive and widespread than anticipated, with almost all counties receiving some number of additional migrants due to SLR induced flooding. We identified two possible channels through which SLR can affect migration. First, the ``direct effect'' shows the amount of people that are forced to migrate away from flooded areas. The ``indirect effect'' consists of changes in the magnitude and pattern of migration across all populated areas due to the impact of SLR. The areas immediately adjacent to coastal counties will experience the most dramatic impacts as many migrants will simply move slightly further inland. These counties are predominantly rural and may not have the infrastructure capacity to serve several times the number of average annual incoming migrants. Increased competition for scarce resources, tighter housing and labor markets and more congested roads and diminished access to amenities can deteriorate the living conditions in the counties that receive more migrants. It is of course possible that the new influx of migrants could stimulate the economy thus creating new opportunities for growth in the local economy. This can be the case, not only if migrants flow from more affluent communities, but also because of increased human capital that can spur innovation and improve productivity. The purpose of our study is not to determine whether or not new influx of people are good or bad for a local economy, but to highlight the range of migration changes that can occur because of sea level rise. The overall socio-economic impacts of climate driven migration, of which our results are the first piece, need to be further studied as more data becomes available. 

In general, we find that previously ``unpopular'' destinations would be more popular solely due to their close proximity to counties that experience ``direct effects''. The East Coast will experience larger effects than the West coast because of the large coastal population centers and shallower coastlines. Existing urban areas will receive the largest magnitudes of migrants, as they represent the most attractive destinations, which will accelerate the existing trends of urbanization. We find that the southeast portion of the United States will experience disproportionately high effects from SLR-driven flooding due to the large vulnerable populations in New Orleans and Miami. These results show that by driving human migration, the impacts of SLR have the potential to be much farther reaching than the coastal areas which they will flood.

We find similar conclusions to previously published estimates of human migration under SLR in the USA~\cite{hauer2017migration} - inland areas immediately adjacent to the coast, and urban areas in the southeast US will observe the largest effects from SLR driven migration. Our method for modeling human migration reveals several notable differences. According to Hauer 2017, the Austin, Texas Core Based Statistical Area is expected to observe the largest effects out of all destinations, with over 800 thousand incoming migrants due to SLR. This result happens because Austin has consistently been one of the fastest growing US cities over the past decade, which is captured and projected by a time series based migration model. Our migration model instead captures the dynamics of human migration between US counties based on population and distance features, and uses this to predict flows between counties without regard to potential short term historic trends. This approach has the benefit of allowing our model to predict flows between pairs of counties for which there are no historical flows, but can also result in potentially underpredicting flows to areas growing at faster-than-average rates. Indeed, in Texas, our results show more incoming migrants to Houston and Dallas - two larger cities closer to affected coastal areas.

Our analysis extends to how ``standard'' migration patterns between unaffected counties can change due to changes in viable destinations. The bottom right panel of Fig \ref{fig:2} shows how the incoming migrant distribution from \textit{unaffected} counties changes in the High 2100 scenario. We observe that migrants from unaffected areas, that would previously move to coastal areas, will especially relocate to larger population centers. The counties surrounding Los Angeles in particular could see tens of thousands of migrants that are not coming from affected areas, but must choose a different location as a result of coastal flooding. 

Because we rely on machine learning techniques, we forfeit the explanatory power of our model in favor of a more accurate prediction. This approach is suitable for the purposes of our research question and conceives a much more flexible methodology to analyze future migration. This black-box approach can be further calibrated as more data on similar temporal and spatial scales for empirical studies to explain migration behaviors becomes available~\cite{piguet2010linking}.

In the meantime, as various aspects of human migration are better understood, especially ones related to environmental pressures, and better models of human migration are created, our flexible analytical framework will easily support new improved implementations and produce more accurate results. Similarly, as SLR flooding estimates are improved - with finer resolution population projections, uncertainty estimates, and models of the potential spatial effects of SLR such as expected flood frequency - the results given by our framework can be refined.

\section*{Acknowledgements}
J.M-C was supported in part by the National Science Foundation of the United States of America (Grant no. 1510510) and the Canada Research Chairs program. B.D. was supported in part by NSF grants CCF-1522054 (COMPUSTNET: Expanding Horizons of Computational Sustainability) and BCS-1638268 (CRISP: Sustainable and Resilient Design of Interdependent Water and Energy Systems at the Infrastructure-Human-Resource Nexus).

\bibliographystyle{ieeetr}
\bibliography{citations}

\end{document}

\maketitle

%--------------------------------------------------------------------
%--------------------------------------------------------------------
%--------------------------------------------------------------------
\section{Code and reproducibility}
We have published open source code for reproducing the results in this paper at \url{https://github.com/calebrob6/migration-slr}. This effort includes scripts for processing the Digital Coast SLR data, reproducing the block group population estimates from Hauer et al. 2016~\cite{hauer2016millions}, training and running the artificial neural network (ANN) migration models described in Robinson and Dilkina~\cite{robinson2018machine}, and creating all results and figures in the paper. We hope that this effort will encourage further study into the effects of sea level rise (SLR) on human migration and in other salient aspects of society.

%--------------------------------------------------------------------
%--------------------------------------------------------------------
%--------------------------------------------------------------------
\subsection{Population projections}
The population projections in this paper are reproduced from the methodology described in Hauer et al. 2016~\cite{hauer2016millions}. This methodology involves modeling the growth (or decline) of housing units, $h^t_i$, for each census block group in the US based on historical data from 1940-2010, then projecting the number of housing units per census block group out to 2100. The population per housing unit, $d_i$, and the group quarters population, $g_i$, for each block group is assumed to stay constant at their 2010 values. Now, the total population for a particular block group, $i$, at some time, $t$, is given as $p^t_i = h^t_i * d_i + g_i$.

We find our estimates of affected county level populations in 2100 under both the medium and high SLR scenarios are similar to those reported in the the Supplementary Information of Hauer et al. 2016. These estimates can be downloaded in the accompanying code repository.

%--------------------------------------------------------------------
%--------------------------------------------------------------------
%--------------------------------------------------------------------
\section{Implementation and results using the Extended Radiation model}
We implement our proposed framework using the extended radiation model~\cite{yang2014limits} as a human migration model instead of the ANN model proposed in the main text. The extended radiation model is a rederivation of the original radiation model~\cite{simini2012universal} under a survival analysis framework that includes the addition of a parameter which controls the influence of scale of the region and the degree of heterogeneity in the distribution of destination locations. Other human mobility models such as the gravity model~\cite{zipf1946p, erlander1990gravity, lenormand2012universal, lenormand2016systematic}, and Schneider's intervening opportunities model~\cite{schneider1959gravity} provide alternate ways to estimate the number of people that migrate between different locations.

The extended radiation model is given in Equation~\ref{equ:extendedRadiationModel}, where the variables are as follows: $P_{ij}$ is the probability that a migrant who leaves zone $i$ will travel to zone $j$, $T_{ij}$ is the number of migrants that travel from zone $i$ to $j$, $m_i$ is the population of zone $i$, $\beta$ is a parameter of the extended radiation model that controls the influence of scale of the region on trips, $d_{ij}$ is the distance between zones $i$ and $j$, and $s_{ij}$ is the population in the circle centered at $i$ with radius $d_{ij}$ (which does not include $m_i$ or $m_j$).

\begin{equation}
\label{equ:extendedRadiationModel}
P_{ij} = \frac{[(m_i + m_j + s_{ij})^\beta - (m_i + s_{ij})^\beta] (m_i^\beta + 1)}{[(m_i + s_{ij})^\beta + 1][(m_i + m_j + s_{ij})^\beta + 1]}, \;\;\;\; i \neq j
\end{equation}

As in the main text, we use two ``versions'' of the extended radiation model to model migrations from flooded areas and unflooded areas separately. We calculate $\beta_C$ and $\beta_S$, the parameters for the models $MIGRATION_C$ and $MIGRATION_S$ respectively, by minimizing the \textit{Common Part of Commuters} (CPC) metric~\cite{lenormand2016systematic} over the same sets of migration data that were used to train the previous ANN model. Here $\beta_C = 0.13$ and $\beta_S = 0.33$. We compare the fit of the ANN and Extended Radiation model in Section \ref{sec:validation}.

\begin{figure*}[!th]
\centering
\includegraphics[width=0.7\linewidth]{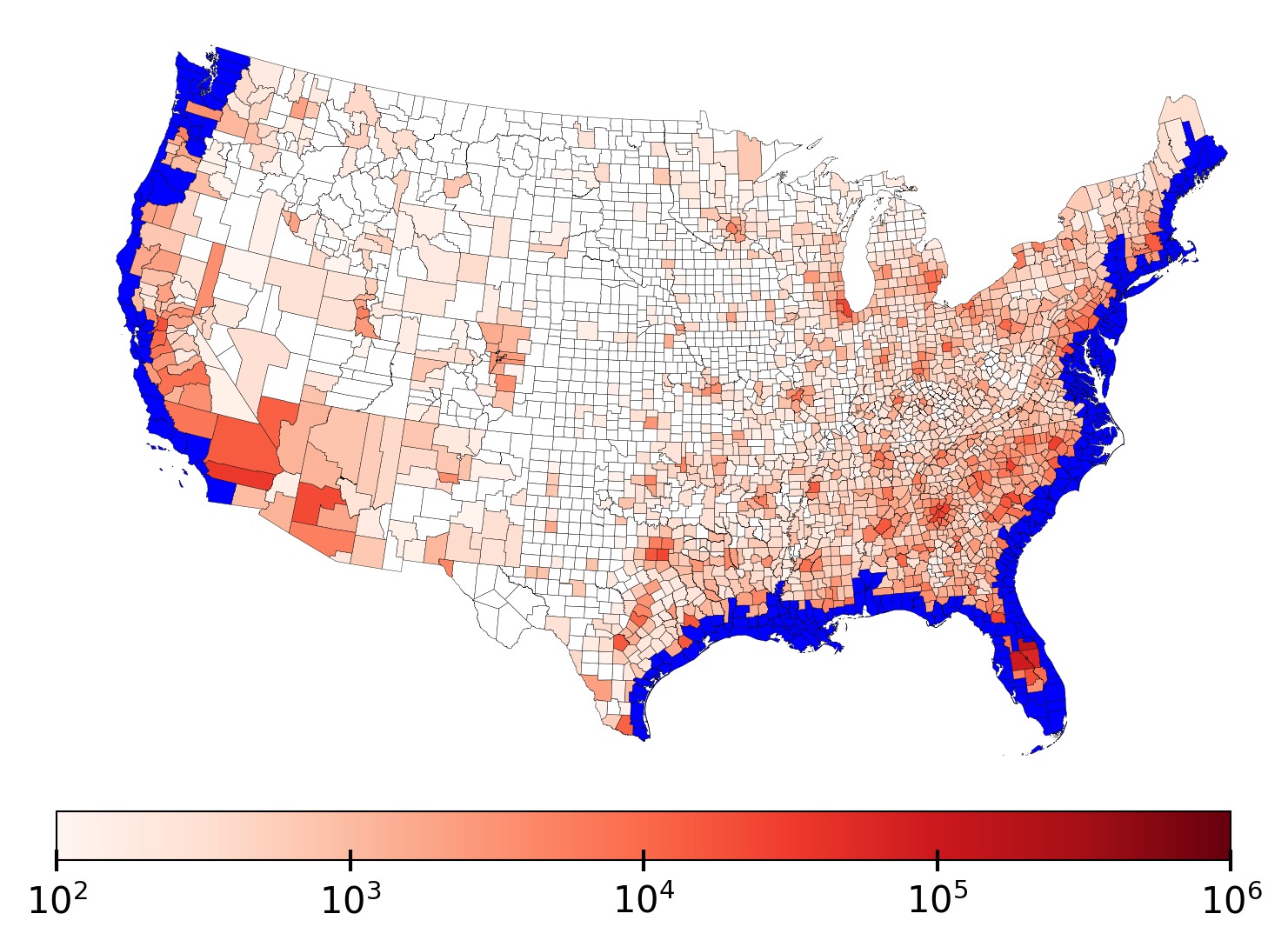}
\includegraphics[width=0.49\linewidth]{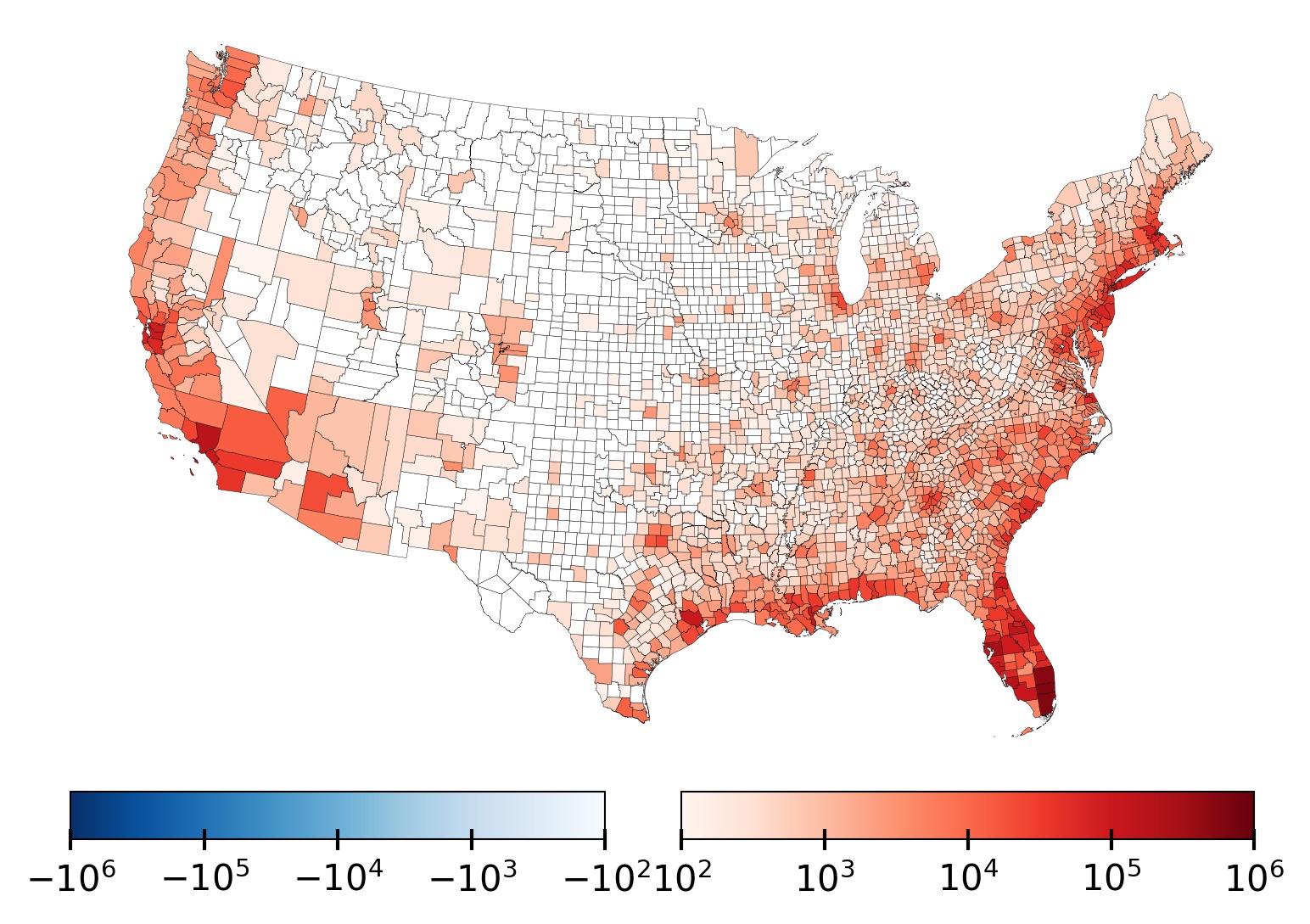}
\includegraphics[width=0.49\linewidth]{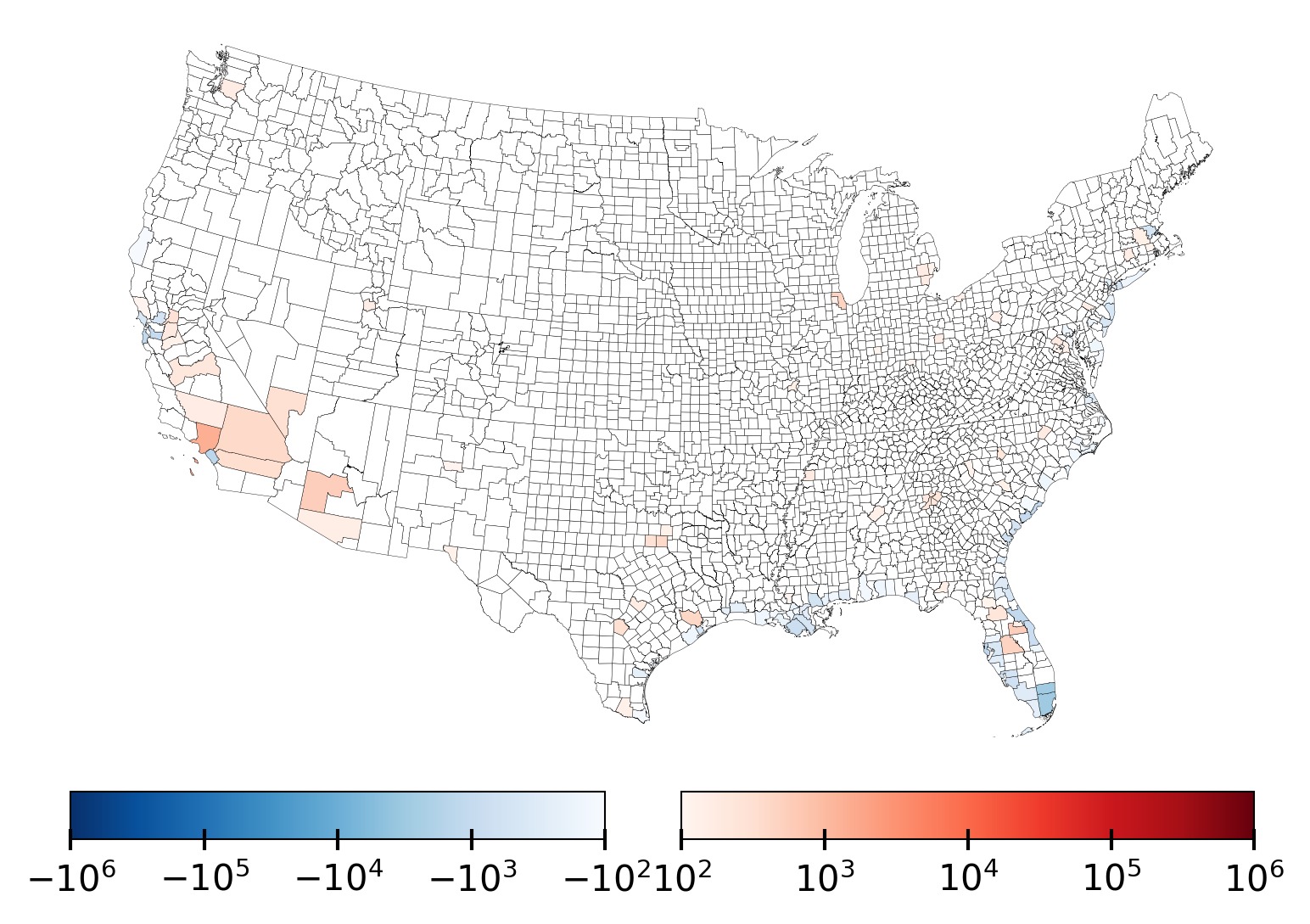}
\caption{\textbf{Extended Radiation Model.} Spatial distribution of the direct and indirect effects of SLR on human migration. The \textbf{top} panel shows all counties that experience flooding under 1.8m of SLR by 2100 in blue and colors the remaining counties based on the number of additional incoming migrants per county that there are in the SLR scenario over the baseline. The \textbf{bottom left} map shows the number of additional incoming migrants per county in the SLR scenario from only flooded counties. The \textbf{bottom right} map shows the number of additional incoming migrants per county in the SLR scenario from only unflooded counties. Color gradients are implemented in a log scale.}
\label{fig:rad_quantities}
\end{figure*}

\begin{figure}[!th]
\centering
\includegraphics[width=0.75\linewidth]{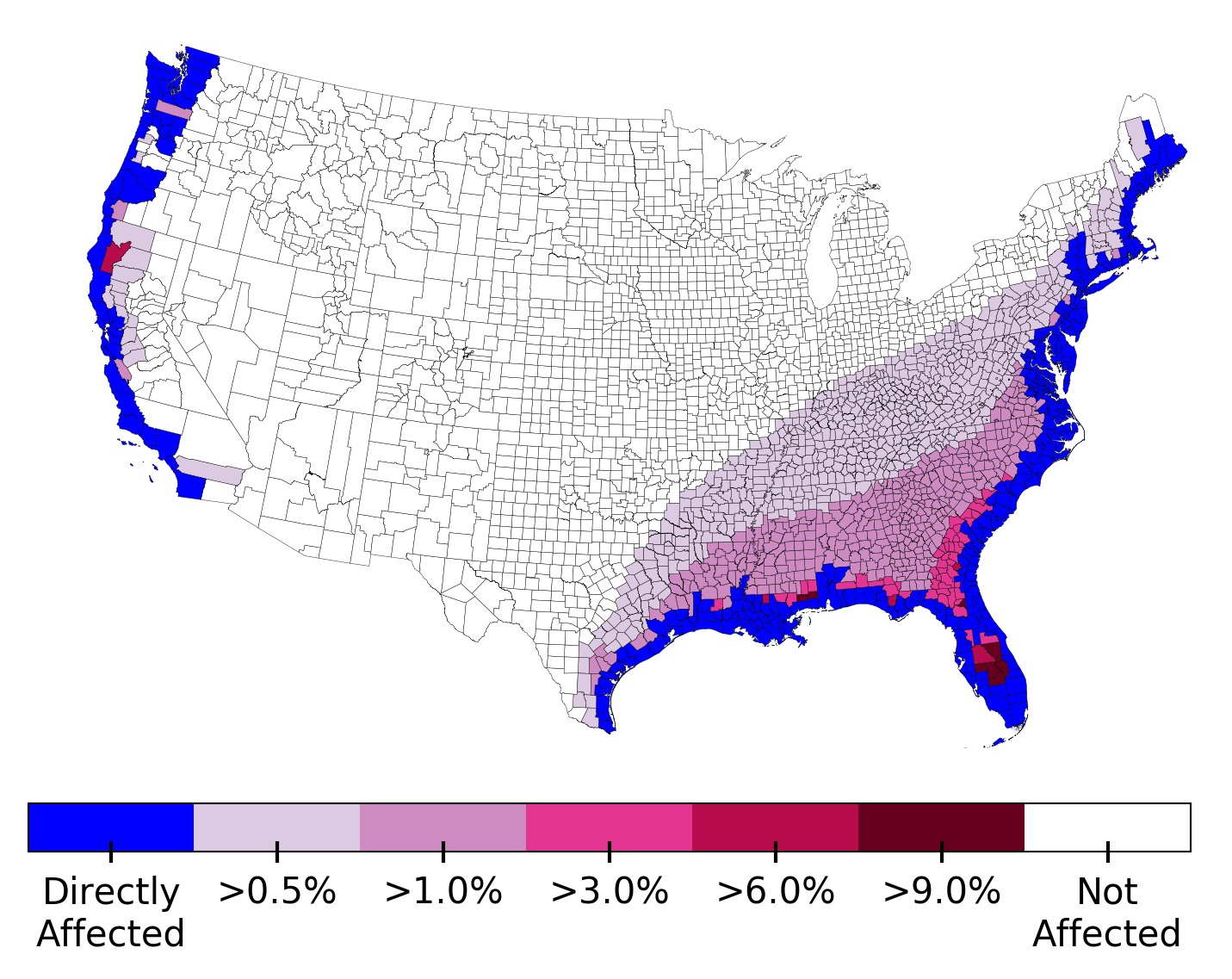}
\includegraphics[width=0.75\linewidth]{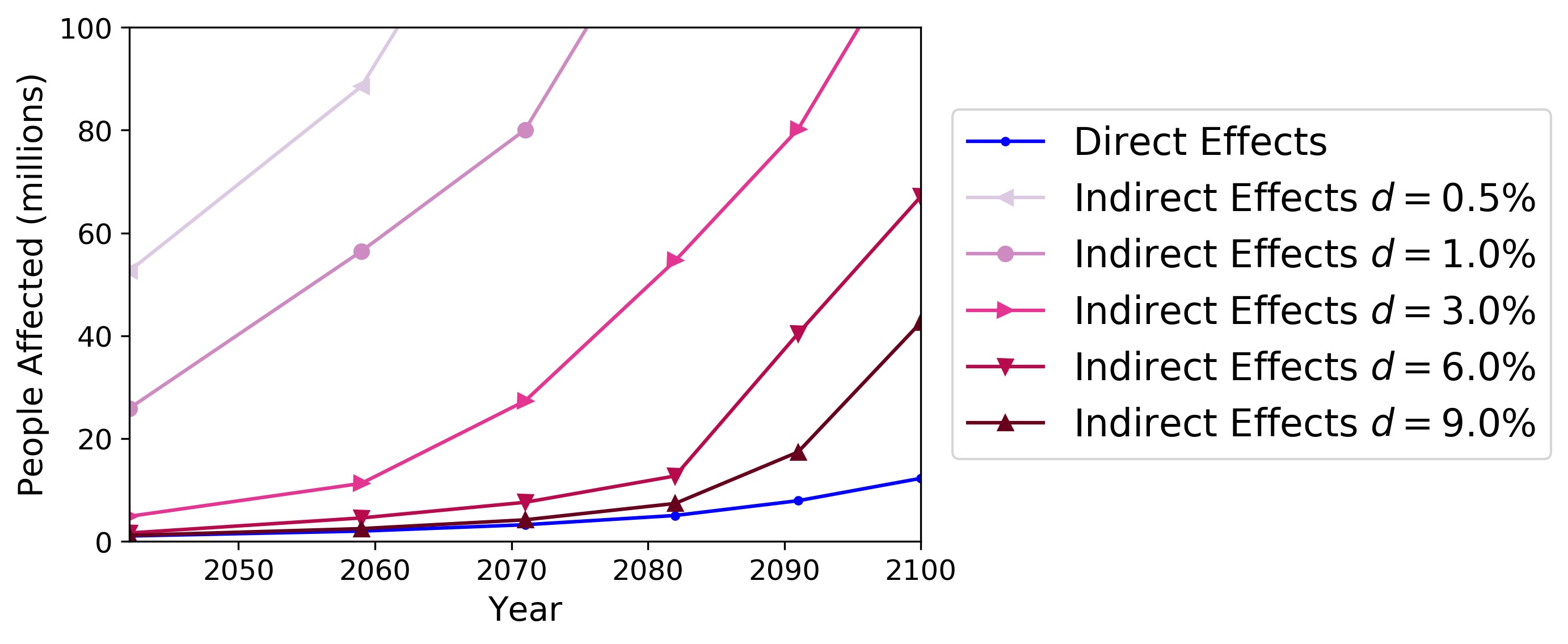}
\caption{\textbf{Extended Radiation Model.} Impacts of SLR due to flooding and human migration for a range of SLR scenarios. In the \textbf{top} panel we show the spatial distribution of counties that are considered indirectly affected at different threshold values of $d$ for the 1.8m SLR case in the southeast portion of the United States. In the \textbf{bottom} panel we show the number of people that are directly and indirectly affected under the same threshold values of $d$ for the entire United States. We show the same impacts from Figure 3 in the main text.}
\label{fig:rad_proportions}
\end{figure}

Figures \ref{fig:rad_quantities} and \ref{fig:rad_proportions} show the results from using these extended radiation models and are in the same format as the results figures from the main text. The general patterns match those from main text (which use an ANN model) - higher concentrations of migrants move to urban areas and counties immediately inland from coastal counties have higher percentages of incoming migrants by population. Figure \ref{fig:rad_proportions} shows the severity of indirect effects degrading with distance from the East coast of the US, while showing that counties that are adjacent to the West coast will not experience similar effects. This result is a combination of the larger displaced populations along the East coast, and the relatively higher density of populations of counties in the Eastern US. The extended radiation model describes the probability of migration between two counties as decaying as a function of the intervening opportunities between them. The intervening opportunities between counties are fewer in the less dense counties of the Western US, meaning longer migrations will be more probable, while shorter migrations will be likely in the more dense Eastern US. Therefore, affected migrants leaving large populations centers along the east coast, such as Miami-Dade in Florida, will diffuse across the available opportunities at a rate roughly proportional to the distance from their origin - giving the observed pattern.

The results given by the extended radiation model predict larger amounts of indirect effects at the higher values of $d$ than the ANN models do, i.e. that there will be larger numbers of people that are indirectly effected by SLR. This is a direct consequence of the diffusion pattern described in the previous paragraph, more population mass is dispersed over larger numbers of rural counties in the southeast US, thus counting the entire populations of these counties as being affected under higher values of $d$.

%--------------------------------------------------------------------
%--------------------------------------------------------------------
%--------------------------------------------------------------------
\section{Migration model training and validation} \label{sec:validation}

Our implementation of the Joint Model relies on ANN models of human migration to estimate migrations from affected and  unaffected counties. We fit these models using historical county-to-county migration data from 2004-2014 from the IRS~\cite{irsData}. In this section, we evaluate the performance of our ANN models against the Extended Radiation~\cite{yang2014limits}, Radiation~\cite{simini2012universal}, and two types of Gravity models~\cite{lenormand2012universal,lenormand2016systematic}. We have three sets of counties: \textbf{all counties} which consists of every county from \textit{all} years of migration data, \textbf{affected counties} which consist of the 7 counties most heavily impacted by Hurricanes Katrina and Rita in 2005 (discussed in the main text), and \textbf{unaffected counties} which consists of the set difference \textbf{all counties} - \textbf{affected counties}. We evaluate each model's average cross-validated performance in three tasks: estimating migrations from \textbf{all counties} to \textbf{all counties},  from \textbf{unaffected counties} to \textbf{unaffected counties}, and from \textbf{affected counties} to \textbf{unaffected counties}. In all of these cross-validation experiments we split on origin counties, i.e. we select a set of training counties and use all migration observations originating in that set of counties (different rows of $\mathbf{T}$) to fit our models, then test on the migration observations originating in the remaining counties. In the \textbf{affected counties} to \textbf{unaffected counties} estimation task we use leave-one-out cross-validation as we only have 7 origin counties, while in the other two tasks we use 5-folds cross-validation. In each cross-validation fold we fit a model for every year of training data and average the results.
This fitting procedure involves: training the ANN models using the parameters/architecture described in~\cite{robinson2018machine}, estimating the single parameter, $\beta$, of the extended radiation or gravity models, and estimating the production function coefficient $\alpha$. We estimate the $\beta$ parameters by maximizing the $CPC$ metric (described below) between the training migration data and modeled migration data. Finally, we estimate the $\alpha$ parameter by calculating the slope of the best fit line through all ($m_i$, $\sum_{j=1}^n T_{ij}$) points, i.e. computing the average fraction of the population of a county that migrates away in a given year. All parameter fitting is done only within a fold's training data, and the average and standard deviations of the $\alpha$ and best $\beta$ parameters for each task is reported in Table \ref{tbl:all_results}.

To measure model performance we record the Common Part of Commuters ($CPC$)~\cite{lenormand2012universal}, Common Part of Commuters distance variant ($CPC_d$)~\cite{lenormand2016systematic}, mean absolute error ($MAE$), and coefficient of determination ($r^2$) between the ground truth migrations, $\mathbf{T}$, and the model estimated migrations, $\mathbf{\hat{T}}$. These metrics are calculated on the origin destination migration matrices, hence we refer to them as ``metrics on full matrix'' in Table \ref{tbl:all_results}.

\begin{description} %
	%----------------------------------------------------------------------------------------------
	%----------------------------------------------------------------------------------------------
	\item[Common Part of Commuters ($CPC$)] This metric directly compares numbers of travelers between the predicted and ground truth matrices. It will be $0$ when the two matrices have no entries in common, and $1$ when they are identical.
	\begin{equation}
	CPC(\mathbf{T},\mathbf{\hat{T}}) = \frac{2 \sum_{i,j=1}^{n} min(T_{ij}, \hat{T}_{ij})}{\sum_{i,j=1}^{n} T_{ij} + \sum_{i,j=1}^{n} \hat{T}_{ij}}
	\end{equation} 
	%----------------------------------------------------------------------------------------------
	%----------------------------------------------------------------------------------------------
	\item[Common Part of Commuters Distance Variant ($CPC_d$)] This metric measures how well a predicted migration matrix recreates trips at the same distances as the ground truth data. In this definition, $N$ is a histogram where a bin $N_k$ contains the number of migrants that travel between $2k-2$ and $2k$ kilometers. It will be $0$ when the two matrices do not have any migrations at the same distance, and $1$ when all fall within the same distances.
	\begin{equation}
	CPC_d(\mathbf{T},\mathbf{\hat{T}}) = \frac{2 \sum_{k=1}^\infty min(N_k, \hat{N}_k)}{\sum_{k=1}^\infty N_k + \sum_{k=1}^\infty \hat{N}_k}
	\end{equation}
	%----------------------------------------------------------------------------------------------
	%----------------------------------------------------------------------------------------------
	\item[Mean absolute error ($MAE$)] This is a standard error measure, the average absolute difference between the predicted and ground truth values. Here, smaller values represent smaller errors in terms of number of migrants.
	\begin{equation}
	MAE(\mathbf{T},\mathbf{\hat{T}}) = \frac{1}{n}\sum_{i,j=1}^{n} (T_{ij} - \hat{T}_{ij})
	\end{equation}
	%----------------------------------------------------------------------------------------------
	%----------------------------------------------------------------------------------------------
	\item[Coefficient of determination ($r^2$)] This score measures the goodness of fit between a set of predictions and the ground truth values. This score ranges from $1$, in a perfect fit, to arbitrarily negative values as a fit becomes worse, and is $0$ when the predictions are equivalent to the expectation of the ground truth values.
	\begin{equation}
	r^2(\mathbf{T},\mathbf{\hat{T}}) = 1 - \frac{\sum_{i,j=1}^n (T_{ij} - \hat{T}_{ij})^2}{\sum_{i,j=1}^n (T_{ij} - \bar{T})^2}
	\end{equation}
	%----------------------------------------------------------------------------------------------
	%----------------------------------------------------------------------------------------------
\end{description}

We also measure the aggregate model performance of predicting the aggregate incoming migrants per county. We calculate the $MAE$ and $r^2$ between $T_{:i} = \sum_{j=1}^n T_{ji}$ and $\hat{T}_{:i} = \sum_{j=1}^n \hat{T}_{ji}$ and call them ``metrics on incoming migrants vector'' in Table \ref{tbl:all_results}.

The results of these experiments are shown in Table \ref{tbl:all_results}. Here, the ANN model outperforms the other models in the large \textbf{unaffected counties} to \textbf{unaffected counties} task and \textbf{all counties} to \textbf{all counties} tasks. In these two tasks the ANN model has a large amount of training data to exploit, and performs best in the matrix $MAE$, matrix $r^2$, and incoming migrants $MAE$ metrics, and second best in the remaining metrics. The gravity model with power law decay performs best in terms of $CPC$ and $CPC_d$ on both tasks, however fails to appropriately model the pairwise flows with a matrix $r^2$ score of near 0. Similarly, the extended radiation model performs slightly better than the ANN model in terms of incoming migrants $r^2$ (and is indeed the only traditional migration model with a positive matrix $r^2$ metric), but performs significantly worse than the ANN model in all other dimensions.

In the \textbf{affected counties} to \textbf{unaffected counties} task, the best performing model is not obvious. Here, the matrix $MAE$ and $r^2$ values are identical to the incoming migrants $MAE$ and $r^2$ values due to the leave-one-out cross validation method - the test set in each split are all possible migrations originating from a single county. We observe that the extended radiation model performs the best in terms of $CPC$ and $r^2$ scores, however with wildly different best parameter values between cross-validation splits (shown in the large standard deviation of the best $\beta$ value). The ANN model is performing best in terms of $CPC_d$ and $MAE$, but poorly on average considering the average $r^2$ score near 0. In the testing splits where the model is performing poorly it has overfit to the longer distance migrations observed in 6 out of the 7 \textbf{affected counties} and is unable to capture the shorter distance migrations from the held out county. As the extended radiation model captures the \textit{general} behavior of human migration (with a single parameter), it does not overfit in this case. The best $\beta$ parameters in the \textbf{affected counties} to \textbf{unaffected counties} task are significantly different than in the other two tasks, showing how the structure of migration may be different under extreme flooding events.

\begin{table}[tp]
\centering
\resizebox{\textwidth}{!}{%
\begin{tabular}{@{}l|r|rrrr|rr@{}}
\toprule
 & \multicolumn{1}{l|}{} & \multicolumn{4}{c|}{\textbf{Metrics on full matrix}} & \multicolumn{2}{c}{\begin{tabular}[c]{@{}c@{}}\textbf{Metrics on incoming} \\ \textbf{migrants vector}\end{tabular}} \\ \midrule
%--------------------------------------------------------------------------------------------
%--------------------------------------------------------------------------------------------
\multicolumn{1}{c|}{\begin{tabular}[c]{@{}c@{}}\textbf{Unaffected to Unaffected}\\ Best $\alpha$ = 0.0325 (0.0017)\end{tabular}} & \multicolumn{1}{c|}{Best $\beta$} & \multicolumn{1}{c}{CPC} & \multicolumn{1}{c}{CPC$_d$} & \multicolumn{1}{c}{MAE} & \multicolumn{1}{c|}{R2} & \multicolumn{1}{c}{MAE} & \multicolumn{1}{c}{R2} \\ \midrule
\multicolumn{1}{l|}{Extended Radiation} & \multicolumn{1}{r|}{\begin{tabular}[c]{@{}r@{}}0.3537\\ (0.0172)\end{tabular}} & \begin{tabular}[c]{@{}r@{}}0.4926\\ (0.0027)\end{tabular} & \begin{tabular}[c]{@{}r@{}}0.7134\\ (0.0436)\end{tabular} & \begin{tabular}[c]{@{}r@{}}1.0495\\ (0.3343)\end{tabular} & \multicolumn{1}{r|}{\begin{tabular}[c]{@{}r@{}}0.1493\\ (0.1027)\end{tabular}} & \begin{tabular}[c]{@{}r@{}}287.7557\\ (84.2073)\end{tabular} & \textbf{\begin{tabular}[c]{@{}r@{}}0.8588\\ (0.0328)\end{tabular}} \\
\multicolumn{1}{l|}{Radiation} & \multicolumn{1}{r|}{n/a} & \begin{tabular}[c]{@{}r@{}}0.4327\\ (0.0109)\end{tabular} & \begin{tabular}[c]{@{}r@{}}0.6426\\ (0.0611)\end{tabular} & \begin{tabular}[c]{@{}r@{}}1.1816\\ (0.3992)\end{tabular} & \multicolumn{1}{r|}{\begin{tabular}[c]{@{}r@{}}-0.6345\\ (0.1548)\end{tabular}} & \begin{tabular}[c]{@{}r@{}}326.2592\\ (89.5871)\end{tabular} & \begin{tabular}[c]{@{}r@{}}0.7773\\ (0.0647)\end{tabular} \\
\multicolumn{1}{l|}{Gravity Exponential Decay} & \multicolumn{1}{r|}{\begin{tabular}[c]{@{}r@{}}0.2124\\ (0.3938)\end{tabular}} & \begin{tabular}[c]{@{}r@{}}0.3058\\ (0.1549)\end{tabular} & \begin{tabular}[c]{@{}r@{}}0.3988\\ (0.2089)\end{tabular} & \begin{tabular}[c]{@{}r@{}}1.4387\\ (0.5281)\end{tabular} & \multicolumn{1}{r|}{\begin{tabular}[c]{@{}r@{}}-3.7014\\ (4.3919)\end{tabular}} & \begin{tabular}[c]{@{}r@{}}373.4170\\ (109.4751)\end{tabular} & \begin{tabular}[c]{@{}r@{}}0.5119\\ (0.1897)\end{tabular} \\
\multicolumn{1}{l|}{Gravity Power Law Decay} & \multicolumn{1}{r|}{\begin{tabular}[c]{@{}r@{}}2.6955\\ (0.0829)\end{tabular}} & \textbf{\begin{tabular}[c]{@{}r@{}}0.5530\\ (0.0244)\end{tabular}} & \textbf{\begin{tabular}[c]{@{}r@{}}0.8000\\ (0.0867)\end{tabular}} & \begin{tabular}[c]{@{}r@{}}0.9421\\ (0.3589)\end{tabular} & \multicolumn{1}{r|}{\begin{tabular}[c]{@{}r@{}}-0.0928\\ (0.8651)\end{tabular}} & \begin{tabular}[c]{@{}r@{}}341.2711\\ (128.2462)\end{tabular} & \begin{tabular}[c]{@{}r@{}}0.5883\\ (0.1862)\end{tabular} \\ \midrule
\multicolumn{1}{l|}{ANN Model} & \multicolumn{1}{r|}{n/a} & \begin{tabular}[c]{@{}r@{}}0.5406\\ (0.0244)\end{tabular} & \begin{tabular}[c]{@{}r@{}}0.7321\\ (0.0500)\end{tabular} & \textbf{\begin{tabular}[c]{@{}r@{}}0.9272\\ (0.3413)\end{tabular}} & \multicolumn{1}{r|}{\textbf{\begin{tabular}[c]{@{}r@{}}0.3687\\ (0.0587)\end{tabular}}} & \textbf{\begin{tabular}[c]{@{}r@{}}265.3275\\ (86.9390)\end{tabular}} & \begin{tabular}[c]{@{}r@{}}0.8435\\ (0.0360)\end{tabular} \\ \midrule
%--------------------------------------------------------------------------------------------
%--------------------------------------------------------------------------------------------
\multicolumn{1}{c|}{\begin{tabular}[c]{@{}c@{}}\textbf{Affected to Unaffected}\\ Best $\alpha$ = 0.1674 (0.0326)\end{tabular}} & \multicolumn{1}{c|}{Best $\beta$} & \multicolumn{1}{c}{CPC} & \multicolumn{1}{c}{CPC$_d$} & \multicolumn{1}{c}{MAE} & \multicolumn{1}{c|}{R2} & \multicolumn{1}{c}{MAE} & \multicolumn{1}{c}{R2} \\ \midrule
\multicolumn{1}{l|}{Extended Radiation} & \multicolumn{1}{r|}{\begin{tabular}[c]{@{}r@{}}0.2403\\ (0.2478)\end{tabular}} & \textbf{\begin{tabular}[c]{@{}r@{}}0.4882\\ (0.0790)\end{tabular}} & \begin{tabular}[c]{@{}r@{}}0.5149\\ (0.0820)\end{tabular} & \begin{tabular}[c]{@{}r@{}}9.2474\\ (12.5452)\end{tabular} & \multicolumn{1}{r|}{\textbf{\begin{tabular}[c]{@{}r@{}}0.5719\\ (0.1355)\end{tabular}}} & - & - \\
\multicolumn{1}{l|}{Radiation} & \multicolumn{1}{r|}{n/a} & \begin{tabular}[c]{@{}r@{}}0.4692\\ (0.1104)\end{tabular} & \begin{tabular}[c]{@{}r@{}}0.4820\\ (0.1087)\end{tabular} & \begin{tabular}[c]{@{}r@{}}11.7799\\ (16.6184)\end{tabular} & \multicolumn{1}{r|}{\begin{tabular}[c]{@{}r@{}}-0.2258\\ (0.7936)\end{tabular}} & - & - \\
\multicolumn{1}{l|}{Gravity Exponential Decay} & \multicolumn{1}{r|}{\begin{tabular}[c]{@{}r@{}}0.0047\\ (0.0002)\end{tabular}} & \begin{tabular}[c]{@{}r@{}}0.3889\\ (0.1446)\end{tabular} & \begin{tabular}[c]{@{}r@{}}0.4184\\ (0.1544)\end{tabular} & \begin{tabular}[c]{@{}r@{}}9.7246\\ (12.4281)\end{tabular} & \multicolumn{1}{r|}{\begin{tabular}[c]{@{}r@{}}0.3068\\ (0.1964)\end{tabular}} & - & - \\
\multicolumn{1}{l|}{Gravity Power Law Decay} & \multicolumn{1}{r|}{\begin{tabular}[c]{@{}r@{}}1.6250\\ (0.0772)\end{tabular}} & \begin{tabular}[c]{@{}r@{}}0.3742\\ (0.1288)\end{tabular} & \begin{tabular}[c]{@{}r@{}}0.4038\\ (0.1414)\end{tabular} & \begin{tabular}[c]{@{}r@{}}10.3996\\ (13.7485)\end{tabular} & \multicolumn{1}{r|}{\begin{tabular}[c]{@{}r@{}}0.3081\\ (0.1584)\end{tabular}} & - & - \\ \midrule
\multicolumn{1}{l|}{ANN Model} & \multicolumn{1}{r|}{n/a} & \begin{tabular}[c]{@{}r@{}}0.4231\\ (0.1275)\end{tabular} & \textbf{\begin{tabular}[c]{@{}r@{}}0.5366\\ (0.2240)\end{tabular}} & \textbf{\begin{tabular}[c]{@{}r@{}}9.0060\\ (10.0530)\end{tabular}} & \multicolumn{1}{r|}{\begin{tabular}[c]{@{}r@{}}-0.0640\\ (0.5563)\end{tabular}}  & - & - \\ \midrule
%--------------------------------------------------------------------------------------------
%--------------------------------------------------------------------------------------------
\multicolumn{1}{c|}{\begin{tabular}[c]{@{}c@{}}\textbf{All to All}\\ Best $\alpha$ = 0.0326 (0.0017)\end{tabular}} & \multicolumn{1}{c|}{Best $\beta$} & \multicolumn{1}{c}{CPC} & \multicolumn{1}{c}{CPC$_d$} & \multicolumn{1}{c}{MAE} & \multicolumn{1}{c|}{R2} & \multicolumn{1}{c}{MAE} & \multicolumn{1}{c}{R2} \\ \midrule
\multicolumn{1}{l|}{Extended Radiation} & \multicolumn{1}{r|}{\begin{tabular}[c]{@{}r@{}}0.3544\\ (0.0169)\end{tabular}} & \begin{tabular}[c]{@{}r@{}}0.4923\\ (0.0031)\end{tabular} & \begin{tabular}[c]{@{}r@{}}0.7145\\ (0.0436)\end{tabular} & \begin{tabular}[c]{@{}r@{}}1.0551\\ (0.3254)\end{tabular} & \multicolumn{1}{r|}{\begin{tabular}[c]{@{}r@{}}0.1467\\ (0.1012)\end{tabular}} & \begin{tabular}[c]{@{}r@{}}290.3686\\ (81.5585)\end{tabular} & \textbf{\begin{tabular}[c]{@{}r@{}}0.8553\\ (0.0323)\end{tabular}} \\
\multicolumn{1}{l|}{Radiation} & \multicolumn{1}{r|}{n/a} & \begin{tabular}[c]{@{}r@{}}0.4326\\ (0.0107)\end{tabular} & \begin{tabular}[c]{@{}r@{}}0.6422\\ (0.0600)\end{tabular} & \begin{tabular}[c]{@{}r@{}}1.1875\\ (0.3895)\end{tabular} & \multicolumn{1}{r|}{\begin{tabular}[c]{@{}r@{}}-0.6341\\ (0.1644)\end{tabular}} & \begin{tabular}[c]{@{}r@{}}329.6186\\ (86.5392)\end{tabular} & \begin{tabular}[c]{@{}r@{}}0.7718\\ (0.0656)\end{tabular} \\
\multicolumn{1}{l|}{Gravity Exponential Decay} & \multicolumn{1}{r|}{\begin{tabular}[c]{@{}r@{}}0.2124\\ (0.3938)\end{tabular}} & \begin{tabular}[c]{@{}r@{}}0.3062\\ (0.1551)\end{tabular} & \begin{tabular}[c]{@{}r@{}}0.3997\\ (0.2091)\end{tabular} & \begin{tabular}[c]{@{}r@{}}1.4451\\ (0.5209)\end{tabular} & \multicolumn{1}{r|}{\begin{tabular}[c]{@{}r@{}}-3.7402\\ (4.4424)\end{tabular}} & \begin{tabular}[c]{@{}r@{}}376.9256\\ (107.9260)\end{tabular} & \begin{tabular}[c]{@{}r@{}}0.5066\\ (0.1895)\end{tabular} \\
\multicolumn{1}{l|}{Gravity Power Law Decay} & \multicolumn{1}{r|}{\begin{tabular}[c]{@{}r@{}}2.6936\\ (0.0816)\end{tabular}} & \textbf{\begin{tabular}[c]{@{}r@{}}0.5522\\ (0.0240)\end{tabular}} & \textbf{\begin{tabular}[c]{@{}r@{}}0.7997\\ (0.0863)\end{tabular}} & \begin{tabular}[c]{@{}r@{}}0.9479\\ (0.3511)\end{tabular} & \multicolumn{1}{r|}{\begin{tabular}[c]{@{}r@{}}-0.1035\\ (0.8609)\end{tabular}} & \begin{tabular}[c]{@{}r@{}}343.9989\\ (125.8610)\end{tabular} & \begin{tabular}[c]{@{}r@{}}0.5822\\ (0.1836)\end{tabular} \\ \midrule
\multicolumn{1}{l|}{ANN Model} & \multicolumn{1}{r|}{n/a} & \begin{tabular}[c]{@{}r@{}}0.5445\\ (0.0237)\end{tabular} & \begin{tabular}[c]{@{}r@{}}0.7367\\ (0.0556)\end{tabular} & \textbf{\begin{tabular}[c]{@{}r@{}}0.9461\\ (0.3451)\end{tabular}} & \multicolumn{1}{r|}{\textbf{\begin{tabular}[c]{@{}r@{}}0.3632\\ (0.0633)\end{tabular}}} & \textbf{\begin{tabular}[c]{@{}r@{}}275.1929\\ (90.9055)\end{tabular}} & \begin{tabular}[c]{@{}r@{}}0.8357\\ (0.0440)\end{tabular} \\ \bottomrule
%--------------------------------------------------------------------------------------------
%--------------------------------------------------------------------------------------------
\end{tabular}%
}
\caption{Comparison of different models in predicting three different classes of migrations. Averaged cross-validation results are shown with standard deviations in parenthesis. In all columns except $MAE$, higher values represent better performance. Best performing number is bolded in each column, independently for each class of migration.}
\label{tbl:all_results}
\end{table}

%--------------------------------------------------------------------
%--------------------------------------------------------------------
%--------------------------------------------------------------------
\section{Effects of modeling climate migrants separately}

\begin{figure}[thp]
    \centering
    \includegraphics[width=0.8\textwidth]{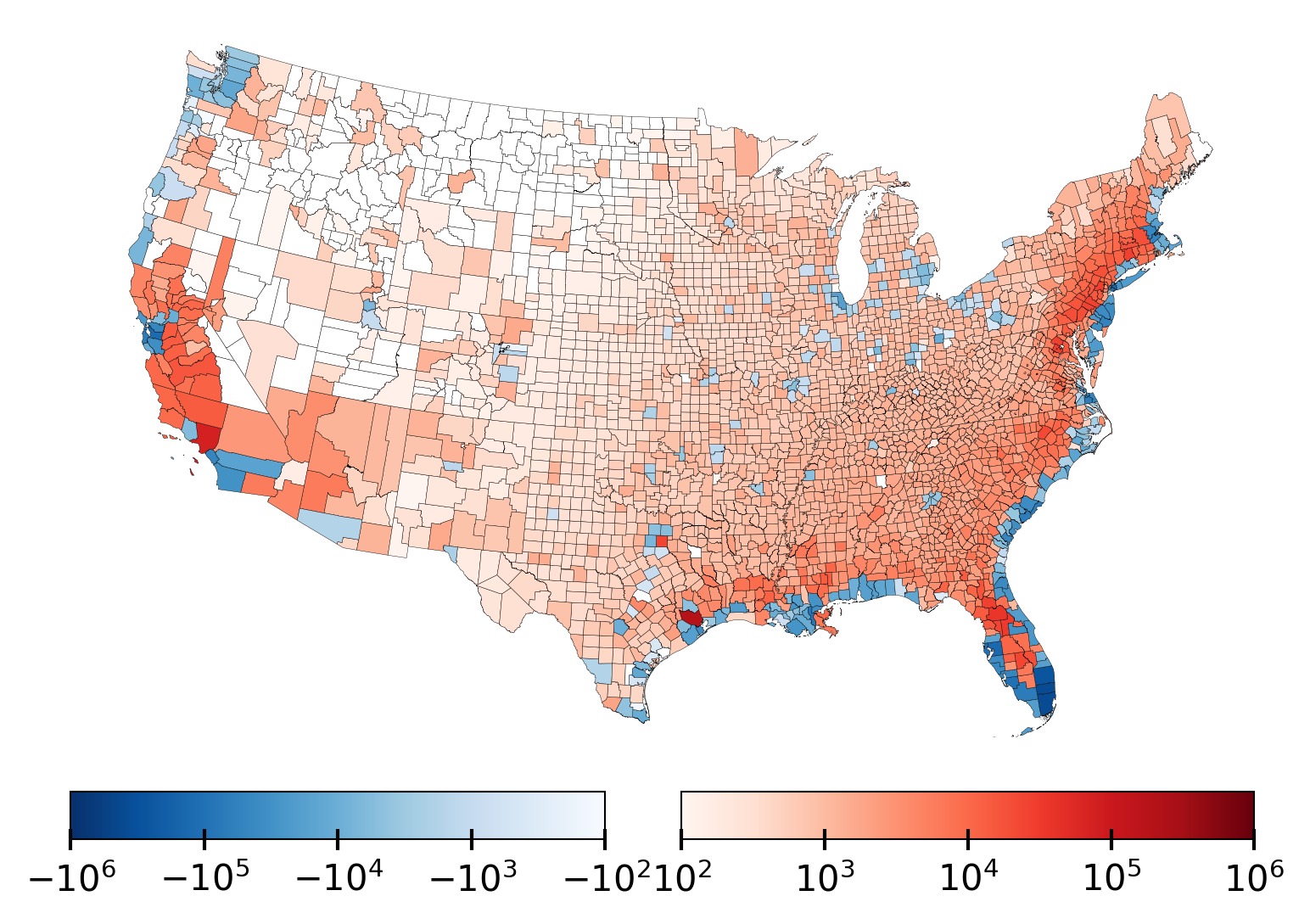}
    \includegraphics[width=0.8\textwidth]{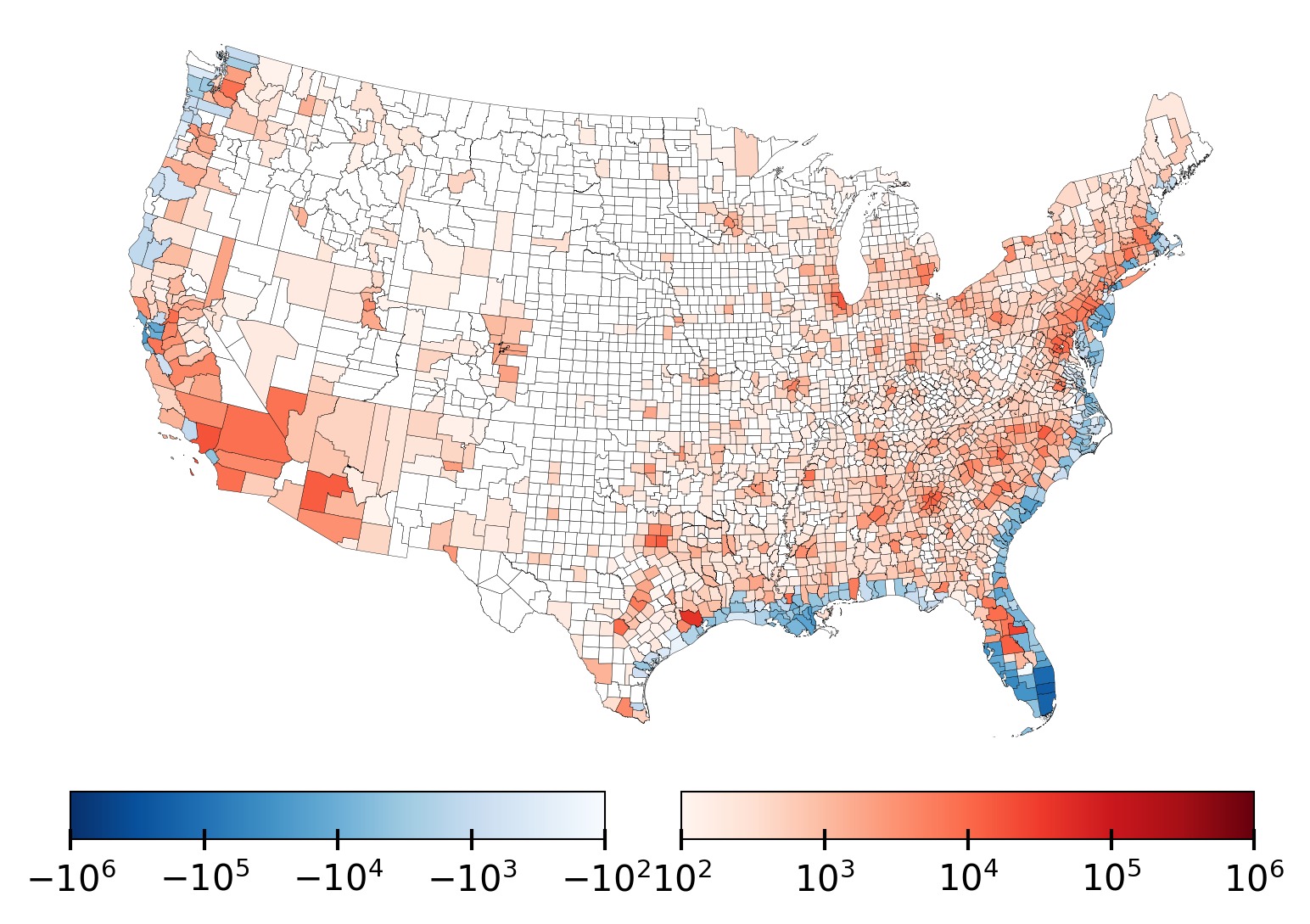}
    \caption{Difference in incoming migrants per county between migrations generated with separately trained models for affected and unaffected migrant behavior and migrations generated with a single model for both. The \textbf{top} panel shows ANN results, and the \textbf{bottom} panel shows Extended radiation model results. All results are for the High 2100 sea level rise/population scenario. In both sets of results the model that \textit{does not} separately capture affected and unaffected migrant behavior predicts more incoming migrants to coastal counties.}
    \label{fig:ablation}
\end{figure}

In the main text we argue that persons living in areas that will be affected by SLR will be exposed to increased climatic pressures and will be forced to migrate elsewhere. Furthermore, we train an ANN model to separately model these migrations based on historic migration patterns from counties that were especially affected by Hurricanes Katrina and Rita. Here, we show the effect that separately modeling these migrations has on our results by simulating the same set of conditions with a single migration model for both types of migrants under both the ANN and Extended Radiation migration models. Figure \ref{fig:ablation} shows the difference in number of incoming migrants per county between the results using separate migration models and the results using a single migration model for the 1.8m SLR scenario. Notably, in both sets of results, when climate driven migrations are \textit{not} modeled separately, then more migrations to coastal areas are predicted. One explanation for this pattern is that the probability of migration for climate driven migrations does not decay as strongly with distance as in standard migrations (e.g. the calibrated scale parameter in the Extended Radiation model is lower for climate migrants, which decreases the impact of intervening opportunities), therefore displaced migrants ``see'' distant population centers as attractive of a destination as nearby coastal destinations (which will usually be more populated than surrounding areas).

One discrepancy between the two sets of results is how, in the Extended radiation model results, separately modeled affected and unaffected migration behavior results in more predicted migrations to urban centers, while the corresponding ANN results show fewer predicted migrations to urban centers. This could also be due to the different emphasis on distance in predicting migration probabilities.

%--------------------------------------------------------------------
%--------------------------------------------------------------------
%--------------------------------------------------------------------
\section{Algorithmic description of the Joint Model}

In the main text we give a high-level outline of our joint migration/SLR modeling framework and provide a detailed implementation using small-area population projections, Digital Coast SLR estimates, and ANN based migration models. Here we describe the steps necessary for implementing the joint model from an algorithmic standpoint.

Our joint model consists of two modules: the sea level rise module and the migration module, which use the \texttt{SLR} and \texttt{MIGRATION} functions respectively. The joint model is independent from the specific implementation of these two functions, and method used to partition features between the affected and unaffected areas in a zone.

A sea level rise model, $SLR(\theta_i, x)$, is a function that takes as input a spatial zone, $\theta_i$, and an amount of SLR, $x$, and outputs the portion of the zone that is flooded under the given amount of SLR. This model should also be used to inform a \textit{splitting} procedure that is needed to partition the zone into flooded and unflooded areas. This procedure can be done with accessory sub-zone resolution data. For example, in our implementation we use spatially explicit flooding data intersected with census block groups in order to partition counties into flooded and unflooded areas.

We describe a migration model, $MIGRATION(\boldsymbol{\theta^S}, \boldsymbol{\theta^D})$, as a function that takes as input a list of `source' zones of length $n$, $\boldsymbol{\theta^S}$, and list of `destination' zones of length $m$, $\boldsymbol{\theta^D}$, and outputs a migration matrix $\mathbf{T} \in \mathbb{N}_0^{n \times m}$, where an entry $T_{ij}$ represents the number of migrants that travel from $\theta_i^S$ to $\theta_j^D$. Our joint model requires two different migration models, $MIGRATION_C$ to handle climate migrations and $MIGRATION_S$, to handle standard migrations. Note that these models \textit{can} be the same in the absence of an appropriate model to handle climate migrations (with the added assumption that climate migrants will behave as standard migrants).

Given a list of zones of length $n$, $\boldsymbol{\theta}$, where each individual zone $\theta_i$ has some number of features associated with it, and an amount of SLR, $x$, our modeling process is as follows:

\begin{itemize}
\item Create two empty lists, $\boldsymbol{\theta}^A$ and $\boldsymbol{\theta}^U$. $\boldsymbol{\theta}^A$ will contain an entry for each zone that represents the `affected' (or flooded) portion of that zone and $\boldsymbol{\theta}^U$ will contain an entry for each `unaffected' portion. Note that $|\boldsymbol{\theta}^A|$ will equal $|\boldsymbol{\theta}^U|$.
\item For each zone, $\theta_i$, calculate $SLR(\theta_i, x)$, and `split' the zone into two separate areas accordingly:
\begin{itemize}
\item If $SLR(\theta_i, x) > 0$, then partition the zone into two areas, $\theta_i^A$ and $\theta_i^U$. $\theta_i^A$ will contain features  of the land that is flooded in the zone, and $\theta_i^U$ will contain the remaining proportions. Append $\theta_i^A$ to $\boldsymbol{\theta}^A$, and append $\theta_i^U$ to $\boldsymbol{\theta}^U$.
\item If $SLR(\theta_i, x) = 0$, then the zone is not affected by SLR. Append $\theta_i$ to $\boldsymbol{\theta}^U$, and append an entry representing a \texttt{null} area to $\boldsymbol{\theta}^A$ in order to keep the lists balanced.
\end{itemize}
\item Calculate the migration matrix $\mathbf{T'} = MIGRATION_C(\boldsymbol{\theta^A}, \boldsymbol{\theta^U})$. This migration matrix will have a size of $n \times n$ and will describe how the people from `flooded' areas will migrate to safe areas.
\item Calculate the migration matrix $\mathbf{T''} = MIGRATION_S(\boldsymbol{\theta^U}, \boldsymbol{\theta^U})$. This migration matrix will also have a size of $n \times n$ and will give how the people from `unaffected' areas will migrate to other `unaffected' areas. Note that we do not consider the case where migrants destinations are in $\boldsymbol{\theta^A}$ because of the assumption that people can not migrate to flooded portions of zones.
\item Output the final migration matrix $\mathbf{T} = \mathbf{T'} + \mathbf{T''}$
\end{itemize}

\bibliographystyle{ieeetr}
\bibliography{citations}